# Oxygen dissociation on the C$_3$N monolayer: A first-principles study

Liang Zhao[a,*], Wenjin Luo[a,*], Zhijing Huang[a], Zihan Yan[a], Hui Jia[a], Wei Pei[a], and Yusong Tu[a,†]

[a]*College of Physical Science and Technology & Microelectronics Industry Research Institute, Yangzhou University, Jiangsu, 225009, China*

*Authors contribute equally to this work
†Corresponding author: ystu@yzu.edu.cn

**Abstract**

The oxygen dissociation and the oxidized structure on the pristine C$_3$N monolayer in exposure to air are the inevitably critical issues for the C$_3$N engineering and surface functionalization yet have not been revealed in detail. Using the first-principles calculations, we have systematically investigated the possible O$_2$ adsorption sites, various O$_2$ dissociation pathways and the oxidized structures. It is demonstrated that the pristine C$_3$N monolayer shows more O$_2$ physisorption sites and exhibits stronger O$_2$ adsorption than the pristine graphene. Among various dissociation pathways, the most preferable one is a two-step process involving an intermediate state with the chemisorbed O$_2$ and the barrier is lower than that on the pristine graphene, indicating that the pristine C$_3$N monolayer is more susceptible to oxidation than the pristine graphene. Furthermore, we found that the most stable oxidized structure is not produced by the most preferable dissociation pathway but generated from a direct dissociation process. These results can be generalized into a wide range of temperatures and pressures using *ab initio* atomistic thermodynamics. Our findings deepen the understanding of the chemical stability of 2D crystalline carbon nitrides under ambient conditions, and could provide insights into the tailoring of the surface chemical structures via doping and oxidation.

## 1. Introduction

The emergence of a wide variety of two-dimensional (2D) materials and the advancement of their promising applications in electronics [1-3], catalysis [4-6], gas sensors [7-9], energy conversion and storage [10-12] inspired by the great success of graphene [13, 14], has been accompanied by an increasing attention on their chemical stability under ambient conditions. Especially, the O$_2$ dissociation involved in the oxidation of 2D materials when exposed to air is the inevitably critical issue in nanomaterial manipulation and device engineering. The presence of chemisorbed oxygen on the pristine surface of 2D materials tunes the bandgaps [15-18], provides active sites for chemical reactions [19] and assists the surface functionalization [20, 21], sometimes is related to the structural degradation and breakdown [22, 23]. Furthermore, we found that the chemisorbed oxygen groups can surprisingly convert carbon-based materials, such as graphene and carbon nanotube, into dynamic covalent materials [24-26] and even induce the self-adaptivity in response to the adsorption of biomolecules [24]. Although the atomic-thick 2D materials have a large surface-volume ratio and are prone to show high surface reactivity, the $O_2$ dissociation on the surface of 2D materials usually depends on the chemical bonding characteristics, thus differs from one to another [14, 27, 28].

Recently, a hole-free and graphene-like 2D honeycomb structure with the stoichiometric formula, $C_3N$, has been successfully synthesized [29, 30] and demonstrated to exhibit ultra-high stiffness [31] and thermal conductivity [32], and excellent electrical properties [33, 34], offering potential applications in photo- or electro-catalysts [35, 36], gas adsorption medium [37] and electrode materials for batteries [38]. Compared with graphene, the $C_3N$ monolayer can be regarded as a nitrogen-doped graphene structure with nitrogen atoms uniformly distributed and all atoms $sp^2$ hybridized. It is expected that the $C_3N$ could exhibit higher surface activity toward small gas molecules like $O_2$ than graphene at ambient conditions due to the nitrogen atoms [39, 40]. However, detailed discussions on the $O_2$ dissociation on the $C_3N$ monolayer and the oxidized structure are still lacking, despite that recent work has pointed out that the $C_3N$ monolayer is chemically inert to $O_2$ molecule [41].

In this work, we performed the first-principles calculations on the oxygen dissociation on a pristine $C_3N$ monolayer. We have systematically explored possible adsorption sites of an oxygen molecule and the atomic oxygen, various oxygen dissociation pathways and the oxidized structures. It is demonstrated that the pristine $C_3N$ monolayer shows more $O_2$ physisorption sites and exhibits the stronger $O_2$ adsorption than the pristine graphene. Among eight $O_2$ dissociation pathways, the most preferable pathway is a two-step process involving an intermediate state with the chemisorbed oxygen molecule where the barrier is lower than that on the pristine graphene, indicating that the pristine $C_3N$ is more susceptible to oxidation than the pristine graphene. These results can also be generalized into a wide range of temperatures and pressures by analyzing the adsorption Gibbs free energy and surface phase diagram according to the *ab initio* atomistic thermodynamics. In addition, we find that the electronic properties can be altered significantly by the oxidized structures with two chemisorbed O atoms, from the metal to the semiconductor with the largest band gap of 0.67 eV. These results deepen the understanding of the chemical stability of 2D crystalline carbon nitrides under ambient conditions, provide insights into the tailoring of surface chemical structures, and may guide the design of high-efficient catalysts via doping and oxidation.

## 2. Model and methods

*2.1 Model of pristine $C_3N$ monolayer*

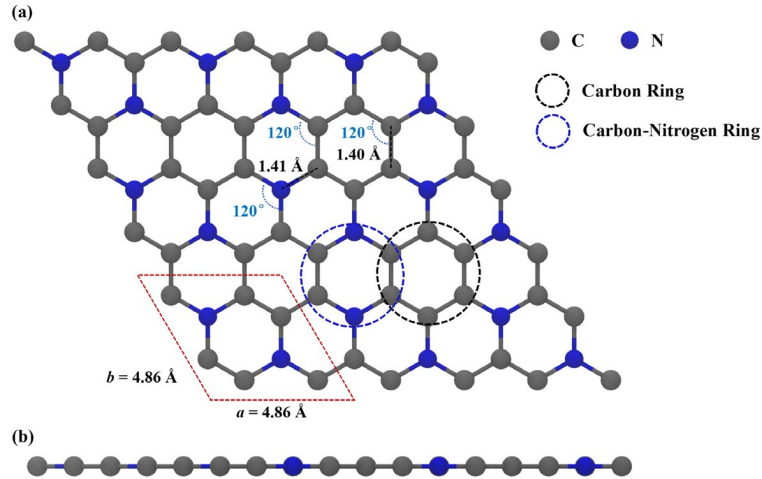

**Fig. 1.** (a) Top and (b) side views of the optimized atomic structure of a pristine C₃N monolayer. The C and N atoms are represented by blue and grey spheres, and the hexagonal primitive unit cell is indicated by the red dashed box. The blue and black dashed circles denote the carbon-nitrogen ring and the carbon ring, respectively.

The pristine $C_3N$ monolayer is modeled within a 3×3×1 supercell, as shown in Fig. 1, including 54 carbon and 18 nitrogen atoms. The optimized lattice constants of $C_3N$, denoted by $a$ and $b$, are $a = b = 4.86$ Å, which are consistent with the experimental value [29]. The average lengths of C-C and C-N bonds are 1.40 Å and 1.41 Å, respectively. The Hirshfeld charges of C and N atoms are +0.01$e$ and -0.03$e$. The calculated values of $E_{C_3N}$, $E_{O_2}$ and $E_O$ are -635.29 eV, -9.85 eV and -1.59 eV and are used throughout this work.

*2.2 Computational details*
All the calculations were performed based on the density functional theory as implemented in the Vienna *Ab*-initio Simulation Package (VASP) [42, 43]. The exchange-correlation interaction is described by the Perdew-Burke-Ernzerhof (PBE) functional within the generalized gradient approximations (GGA) [44], and the interactions between valence electrons and ion-core are described by the projected augmented wave (PAW) method [45, 46]. Grimme's DFT-D3 method is employed to describe the nonlocal dispersive interactions [47]. The energy cutoff of plane wave is 450 eV, and the convergence criteria are 1×10⁻⁵ eV and 0.02 eV/Å for the energy and the force, respectively. The vacuum layer along the *z*-axis direction is 20 Å to avoid the interaction between periodic images. The Monkhorst-Pack scheme is used to sample the *k*-point [48] mesh in the Brillouin zone. For the structure optimization, the 3×3×1 *k*-point mesh is used. For the density of states (DOS) calculations, we use a larger 12×12×1 *k*-point mesh to ensure the accuracy. The size of 3×3×1 supercell containing 54 carbon, 18 nitrogen and 2 oxygen atoms is used in all the calculations, and the system is fully relaxed until the convergence criteria are satisfied. The climbing image-nudged elastic-band (CI-NEB) method [49] is used to search the O₂ molecule dissociation pathway and estimate the reaction barrier connecting the optimized physisorbed and chemisorbed states of C₃N monolayer. All the transition states (TS) are confirmed by the single imaginary frequency output from the default frequency calculation algorithm.

*2.3. Quantitative analyses*

(I) The adsorption energy, $E_{ad}$, is defined by
$$E_{ad} = E_{adsorbate@C_3N} - E_{adsorbate} - E_{C_3N}, \quad (1)$$
where $E_{adsorbate@C_3N}$ is the total energy of the C$_3$N monolayer with the adsorbate (O$_2$ molecule or atomic O), $E_{C_3N}$ is the energy of a pristine C$_3$N monolayer and $E_{adsorbate}$ is the energy of an adsorbate.

(II) The cohesive energy, $E_{coh}$ is defined by [27]
$$E_{coh} = \frac{n_C E_C + n_N E_N - E_{C_3N}}{n_C + n_N}, \quad (2)$$
where $n_C$ and $n_N$ are the numbers of carbon and nitrogen atoms, $E_C$ and $E_N$ represent the energies of atomic carbon (C) and atomic nitrogen (N), respectively.

(III) The O$_2$ dissociation on the C$_3$N monolayer is described by the reaction
$$O_2 + C_3N \rightarrow 2O@C_3N.$$
According to the method developed by Nørskov et al.[50], the Gibbs free energy change of this reaction at a certain temperature $T$ can be estimated by [51, 52]
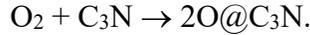
$$\Delta E = E_{2O@C_3N} - E_{O_2} - E_{C_3N}, \quad (3)$$
where $E_{2O@C_3N}$, $E_{O_2}$ and $E_{C_3N}$ are the energies of the absorbed system, an isolated O$_2$ and the pristine C$_3$N monolayer, respectively. $\Delta E_{ZPE}$ is the zero-point energy difference of an O$_2$ molecule between in the gas phase and in the adsorbate state obtained from vibrational frequency calculation. $\Delta S$ is the entropy change of an O$_2$ molecule due to the adsorption.

(IV) The thermodynamical stability of oxidized states at the experimental conditions can be determined by the *ab initio* atomistic thermodynamics. This approach is formulated by Reuter and Scheffler [53-55], and has been an effective tool to explore the interfacial oxidization [56-59]. For the oxidized state of C$_3$N monolayer, the adsorption Gibbs free energy of two chemisorbed O atoms onto the C$_3$N monolayer at the temperature $T$ and O$_2$ partial pressure $p_{O_2}$ is given by $\Delta G^{ad}(T, p_{O_2})$, which is expressed as [60]
$$\Delta G^{ad}(T, p_{O_2}) = G_{2O@C_3N}(T, p_{O_2}) - G_{C_3N}(T, p_{O_2}) - \mu_{O_2}(T, p_{O_2}), \quad (4)$$
where $G_{2O@C_3N}(T, p_{O_2})$ and $G_{C_3N}(T, p_{O_2})$ are the Gibbs free energies of the adsorbed C$_3$N monolayer, the bare C$_3$N monolayer and $\mu_{O_2}(T, p_{O_2})$ is the chemical potential of an O$_2$ molecule. The $\mu_{O_2}(T, p_{O_2})$ is in the following form
$$\mu_{O_2}(T, p_{O_2}) = E_{O_2} + E_{O_2}^{ZPE} + 2\Delta\mu_O(T, p_{O_2}), \quad (5)$$
where $E_{O_2}$, $E_{O_2}^{ZPE}$ are the energy and zero point vibration energy of an O$_2$ molecule. Using the approximation of neglecting the Gibbs free energies of vibrational contribution to the adsorption and configurational entropy in the solid phase, $\Delta G(T, p_{O_2})$ is transformed into the equation [55]

$$\Delta G^{ad}(T, p_{O_2}) \approx E_{ad} - 2\Delta\mu_O(T, p_{O_2}), \quad (6)$$

where $E_{ad} = E_{2O@C_3N} - E_{C_3N} - E_{O_2}$. The $p_{O_2}$-$T$ profile for a given $\Delta\mu_O(T,p)$ can be plotted through [55]

$$\Delta\mu_O(T, p_{O_2}) = -\frac{1}{2}k_B T \left\{ ln\left[\left(\frac{2\pi m}{h^2}\right)^{3/2} \frac{(k_B T)^{5/2}}{p_{O_2}}\right] + ln\left(\frac{k_B T}{\sigma^{sym} B_0}\right) - ln\left[1 - exp\left(-\frac{\hbar\omega_0}{k_B T}\right)\right] + ln(I^{spin}) \right\}, \quad (7)$$

where the values of parameters $\sigma^{sym} = 2$, $I^{spin} = 3$, $B_0 = 0.18\ meV$ and $\hbar\omega_0 = 196\ meV$. The first term in the curly bracket is from the translational free energy of an $O_2$ molecule together with the contribution of $p_{O_2} V$ term. The remaining terms are from the rotational, vibrational and nuclear free energy contributions.

## 3. Results and discussions

*3.1 Physisorption of an oxygen molecule*

The adsorption of an $O_2$ molecule is usually considered to be the first step of $O_2$ dissociation on the surface of nanomaterials. Hence, we search for the favorable physisorption configuration for an $O_2$ molecule on the pristine $C_3N$ monolayer. The adsorption configuration is obtained by initially placing the $O_2$ molecule at the height of about 3.0 Å from the $C_3N$ monolayer and orienting the $O_2$ molecule parallel, vertical or with a tilt angle to the basal plane. As shown in Fig. 2, eight stable adsorption configurations have been found, and can be classified into two types: (I) adsorption sites on the six-membered carbon ring (Figs. (a-d)) and (II) the ones on the six-membered carbon-nitrogen ring (Figs. (e-j)). In the optimized adsorption configurations, the $O_2$ molecule is almost adsorbed in parallel to the basal plane with the height of ~ 3.0 Å and $E_{ad}$ ~ -0.35 eV, consistent with the values in Refs. [37] and [61] which indicate the physisorption. The O-O bond length is 1.27 Å, larger than the 1.23 Å for the isolated $O_2$ molecule, an indication of the weak attraction between the $O_2$ molecule and the $C_3N$ monolayer. The orientation of stable physisorption configurations shows some symmetries: on the carbon ring, the $O_2$ molecule prefers to be adsorbed on the (a) *para*-carbon, (b) *ortho*-carbon, (c) *meta*-carbon and (d) hollow sites; on the carbon-nitrogen ring, the $O_2$ molecule is more easily captured on the (e) *ortho*-carbon-nitrogen, (f) *para*-nitrogen, (g) hollow, (h) *para*-carbon, (i) *meta*-carbon and (j) *meta*-carbon-nitrogen sites. Compared with the $O_2$ adsorption height of ~ 3.0 Å and $E_{ad}$ ~ -0.1 eV on the pristine graphene [40, 62], the $O_2$ molecule is more preferably physisorbed onto the $C_3N$ monolayer and the stable adsorption configurations are enriched due to the presence of N atoms.

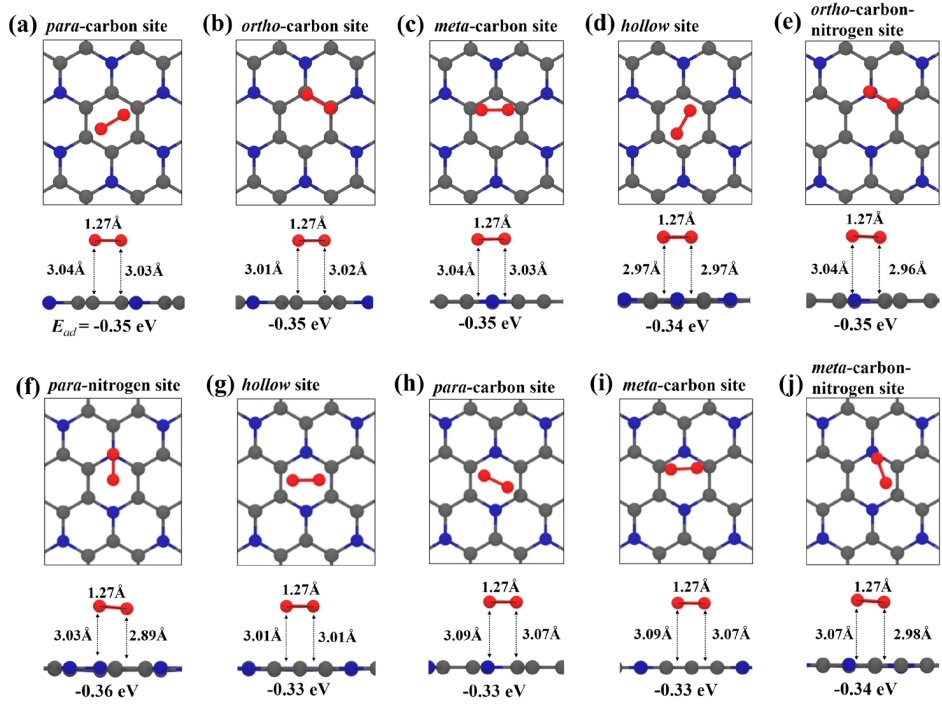

**Fig. 2.** Top and side views of optimized physisorption configurations of O₂ molecule on the pristine C₃N monolayer: (a) *para*-carbon site, (b) *ortho*-carbon site, (c) *meta*-carbon site and (d) hollow site on the six-membered carbon ring, and (e) *ortho*-carbon-nitrogen site, (f) *para*-nitrogen site, (g) hollow site, (h) *para*-carbon site, (i) *meta*-carbon site and (j) *meta*-carbon-nitrogen site on the six-membered carbon-nitrogen ring. The arrows and numbers denote the heights of two oxygen atoms with respect to the basal plane. $E_{ad} = E_{O_2@C_3N} - E_{O_2} - E_{C_3N}$, representing the adsorption energy of an O₂ molecule.

*3.2 Chemisorption of an oxygen atom*

We further explore the chemisorption site for an O atom on the pristine C₃N monolayer as the O₂ molecule will dissociate into two chemisorbed O atoms. As shown in Fig. 3, four possible chemisorption sites are examined, including the top sites of a carbon atom ($T_C$) and a nitrogen atom ($T_N$), the bridge sites of C-C bond ($B_{C-C}$) and C-N bond ($B_{C-N}$). Interestingly, we find that the dangling C-O bond at the $T_C$ site and the dangling N-O bond at the $T_N$ site are energetically stable with the adsorption energies of -4.10 eV and -2.47 eV. The O atom can stay at the $B_{C-C}$ site in the form of C-O-C (epoxy) rather than staying at the $B_{C-N}$ site, which is not shown in Fig. 3. Hirshfeld charge analysis shows that the O at the $T_C$, $B_{CC}$ and $T_N$ sites gains the charge of 0.36*e*, 0.19*e* and 0.33*e*, and the bonded atoms lose the charge of 0.07*e* (carbon atom), 0.08*e* (carbon atom) and 0.11*e* (nitrogen atom), respectively. Compared with the charge loss of 0.01*e* on the C atom and charge gaining of 0.03*e* on the N atom in the pristine C₃N monolayer, the chemisorption of an O atom induces the charge transfer on the bonded atoms. On the pristine graphene, an O atom can only stays at the $B_{C-C}$ site with the $E_{ad}$ ~ -2.32 eV [63], higher than that of -3.61 eV on the C₃N monolayer, indicating that the epoxy configuration is more stable on the C₃N monolayer.

The presence of N atoms enriches the variety of chemical bonds on the surface of $C_3N$ monolayer, such as the dangling C-O and N-O bonds, which are thermodynamically unstable on the pristine graphene.

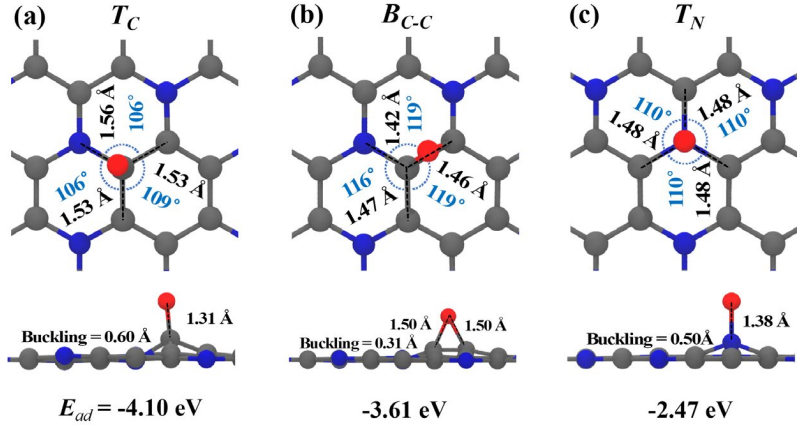

**Fig. 3.** The optimized structures of an O atom chemisorbed on the (a) top site of a carbon atom, (b) bridge site of C-C bond and (c) top site of a nitrogen atom. $T_C$, $B_{C-C}$ and $T_N$ are the denotations for these adsorption sites. The numbers in the top view denote the lengths of bonds between the O-bonded atom and its three covalent atoms, and the angles between bonds are shown in blue numbers. The value of buckling is measured by the height of the O-bonded atom with respect to the plane of the $C_3N$ monolayer. The adsorption energy is defined by $E_{ad} = E_{O@C_3N} - E_O - E_{C_3N}$, representing the adsorption energy of an isolated O atom.

The dangling O atom at the $T_C$ and $T_N$ sites and the epoxy at the $B_{C-C}$ site cause the carbon atom to move out of the plane, inducing the local structural deformation compared to the pristine $C_3N$ monolayer. At the $T_C$ site (Fig. 3(a)), a buckling of 0.60 Å is formed, with two equally C-C bonds elongated from 1.40 Å to 1.53 Å and a C-N bonds elongated from 1.41 Å to 1.56 Å. The dangling C-O bond of 1.31 Å is not strictly pointing along the normal and tilts slightly toward the N atom with the O-C-N angle of 111.8°. Both the geometric configuration and negative adsorption energy of the atomic O at the $T_C$ site are close to the descriptions reported in Refs. [64] and [65]. A more symmetric configuration can be seen at the $T_N$ site (Fig. 3(c)). The dangling N-O bond is along the normal and the buckling is 0.50 Å. The three C-C bonds are all equally elongated to 1.48 Å, with the formation of three equal angles of 110°. At the $B_{C-C}$ site (Fig. 3(b)), two carbon atoms bonded to the O are pulled out to the height of 0.31 Å. The existence of stable dangling C-O or N-O bonds and epoxy enriches the interfacial structures of the $C_3N$ monolayer and may lead to various dissociation pathways of $O_2$ molecule.

*3.3. The oxygen molecule dissociation pathways*

The cohesive energy, $E_{coh}$, is usually used to measure the strength of chemical bonds within the 2D materials and quantify the oxidation resistance. We find that $E_{coh}$ of the pristine $C_3N$ monolayer is 6.99 eV/atom, close to that of graphene and *h*-BN (~ 7 eV/atom) [66-69] and larger than that of germanene and phosphorene (~ 4 eV/atom) [70-72], which indicates that the pristine $C_3N$ monolayer is relatively inert to $O_2$ and exhibits decent oxidation resistance at ambient conditions.

In order to quantify the $O_2$ dissociation process in detail, we focus on the possible dissociation pathways of $O_2$ starting from all the stable physisorption configurations on the pristine $C_3N$ monolayer. As the physisorption of an $O_2$ molecule can be classified into two groups, we will discuss the $O_2$ dissociation on the carbon and carbon-nitrogen rings of the $C_3N$ monolayer individually. In the following discussions, only the $O_2$ or two O atoms, and the adsorbed carbon or carbon-nitrogen ring are shown for clarity.

*3.3.1. $O_2$ dissociation on the carbon ring*

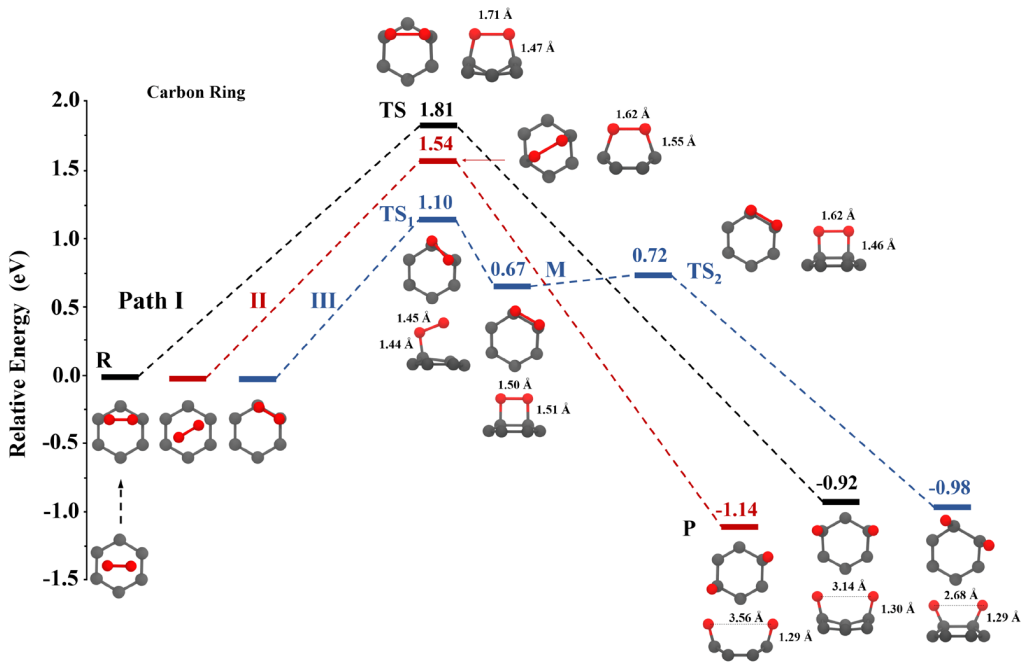

**Fig. 4.** $O_2$ dissociation pathways and optimized configurations on the carbon ring of the $C_3N$ monolayer. Notations: reactant (R), intermediate (M), transition state (TS) and product (P). The energy of R is shifted to zero for an easy comparison and the number on the energy level denotes the relative energy. The relative energy of P state also represents the dissociation energy for the $O_2$ dissociation process. The black numbers on the side view of configurations are the lengths of O-O and C-O bonds, respectively. The black dashed arrow denotes the transformation process between two physisorbed configurations. In all these configurations of R, M, TS and P, only $O_2$ or two O atoms, and the adsorbed carbon ring are shown for clarity.

Paths I and II are one-step dissociation processes where the reactant (R) and product (P) are connected by only one transition state (TS), as shown in Fig. 4, and the dissociation barriers are 1.81 and 1.54 eV, respectively. Path I starts with the $O_2$ physisorbed along the direction of *meta*-carbon atoms of the carbon ring of $C_3N$ monolayer. Then, $O_2$ approaches the $C_3N$ monolayer and two O atoms directly bond to the *meta*-carbon atoms, as indicated by the elongation of O-O distance from 1.27 Å to 1.71 Å and the formation of two C-O bonds of 1.47 Å in TS. Finally, the $O_2$ molecule dissociates into two dangling C-O bonds with the length of 1.30 Å, resulting in the decrease of relative energy from 1.81 eV of TS to -0.92 eV of P state. We consider Path I to be a direct dissociation process as there is only one transition state connecting the R state and P state. Along Path II, starting from the physisorption of $O_2$ on the *para*-carbon site, O-O bond is elongated and two O atoms are bonded to the *para*-carbon atoms. With the increase of O-O bond length to 3.56 Å, $O_2$ molecule dissociates into two dangling C-O bonds with the length of 1.29 Å. We also find that the physisorption configuration of $O_2$ on the hollow site (see Fig. 2(d)), can first shift to that on the *meta*-carbon site via the translation of $O_2$ by overcoming a barrier of 0.62 eV (see PS. 1 in Supplementary Information (SI)).

Different from Paths I and II where the $O_2$ molecule directly dissociates into two chemisorbed O atoms on the $C_3N$ monolayer from the physisorbed configurations, Path III is an indirect and a two-step process involving the chemisorption of $O_2$ as an intermediate for dissociation and exhibits a relatively low barrier of 1.10 eV. Starting from a parallel orientation to the direction of *ortho*-carbon atoms (C-C bond) in R state, the $O_2$ molecule approaches the $C_3N$ monolayer from the adsorption height of about 3 Å by bonding an O atom to the C atom, with the formation of a C-O bond of 1.44 Å in $TS_1$. The O-O bond is tilted with an angle of 101.3° to the C-O bond and elongated from 1.27 Å in R to 1.45 Å. The energy barrier is 1.10 eV, lower than that for Path I and II. As the increase of O-O bond length to 1.50 Å and the formation of two C-O bonds of 1.51 Å, a chemisorbed intermediate (M) is reached and the energy is reduced to 0.67 eV. This step is consistent with that reported in Ma's work [65], where the barrier is 1.04 eV and the energy of M state is 0.69 eV with respect to the R state. In the next step, by passing a low barrier of 0.05 eV, the $O_2$ molecule dissociates into two dangling C-O bonds and the distance between two O atoms reaches the maximum of 2.68 Å. This two-step dissociation of $O_2$ is also found on the pristine graphene where the barrier from the physisorbed state on the *para*-carbon site to the M state is 2.0 eV [73], indicating that the pristine $C_3N$ monolayer shows better catalytic properties on $O_2$ dissociation than on the pristine graphene.

**Table 1**
Calculated total energy ($E_t$), adsorption energy of two chemisorbed O atoms ($E_{ad} = E_{2O@C_3N} - 2E_o - E_{C_3N}$), Hirshfeld charges of two O atoms ($q_O$) and two O-bonded carbon atoms ($q_C$), distances between two dangling O atoms ($d_{OO}$), distances between two O-bonded carbon atoms ($d_{CC}$) of P states with two chemisorbed dangling O atoms on the carbon ring of $C_3N$ monolayer, and the imaginary frequency ($\omega$) of the related TS.

| Sites in P state | $E_t$/eV | $E_{ad}$/eV | $q_O$/e | $q_C$/e | $d_{OO}$/Å | $d_{CC}$/Å | $\omega$/cm$^{-1}$ |
|---|---|---|---|---|---|---|---|
| *meta*-carbon (Path I) | -646.42 | -7.94 | -0.33 -0.33 | +0.07 +0.07 | 3.14 | 2.60 | 1030 |
| *para*-carbon (Path II) | -646.63 | -8.16 | -0.34 -0.34 | +0.08 +0.08 | 3.56 | 2.88 | 613 |
| *ortho*-carbon (Path III) | -646.47 | -8.00 | -0.32 -0.32 | +0.07 +0.07 | 2.68 | 1.63 | 255 (TS$_1$) 697 (TS$_2$) |

The relative energies of P states are negative, indicating that the $O_2$ dissociation processes along Paths I, II and III on the carbon ring of the $C_3N$ monolayer are energetically favorable and exothermic. The adsorption energies $E_{ad}$ for P states with the two dangling C-O bonds are -7.94 eV (Path I, *meta*-carbon site), -8.16 eV (Path II, *para*-carbon site) and -8.00 eV (Path III, *ortho*-carbon site), which are all higher than two times of $E_{ad}$ = -4.10 eV of state with a single dangling C-O bond (see Fig. 3(a)). This is mainly due to the electrostatic repulsion between two negatively charged O atoms of -0.33 *e*, -0.34 *e* and -0.32 *e* for P states in Paths I, II and III. The O…O distances for P states are 3.14 Å, 3.56 Å and 2.68 Å which are larger than the corresponding C…C distances of 2.60 Å for *meta*-carbon atoms, 2.88 Å for *para*-carbon atoms and 1.63 Å for *ortho*-carbon atoms, also an indication of repulsion between two dangling O atoms. In addition, the buckling and the local structure deformation around the bonded carbon or nitrogen atoms can be seen in side views of TS and P states in Fig. 4.

*3.3.2. $O_2$ dissociation on the carbon-nitrogen ring*

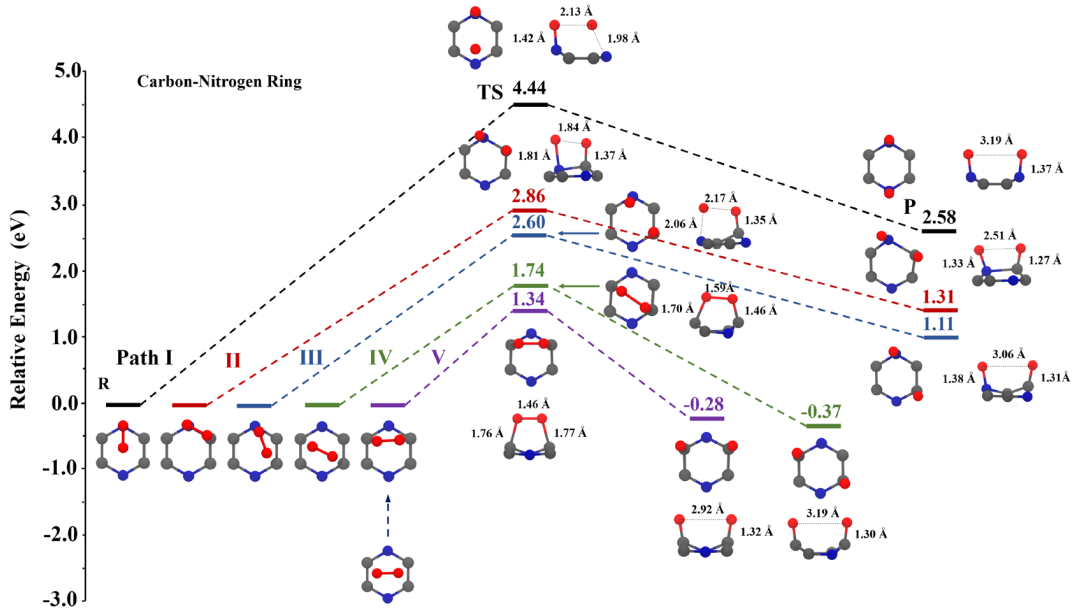

**Fig. 5.** $O_2$ dissociation pathways and optimized configurations on the carbon-nitrogen ring of the $C_3N$ monolayer. Notations, arrows are same to those in Fig. 4. The black values on the side views of configurations denote the lengths of O-O, C-O and N-O bonds. The blue dashed arrow denotes the transformation process between two physisorbed configurations. In all these configurations of R, TS and P, only $O_2$ or two O atoms, and the adsorbed carbon-nitrogen ring are shown for clarity.

Paths I-V on the carbon-nitrogen ring are all one-step dissociation processes with the barriers of 4.44 eV, 2.86 eV, 2.60 eV, 1.74 eV and 1.34 eV, and three of these values are remarkably larger than those on the carbon ring, an indication of more difficult $O_2$ dissociation. Along Path I, an O atom starts to get close to an N atom and the O-N distance decreases from about 3 Å to 1.42 Å. The O-O distance is elongated from 1.27 Å to 2.13 Å meantime another O atom approaches the *para*-nitrogen atom with the O-N distance of 1.98 Å. Next, two dangling N-O bonds of 1.37 Å are formed in the P state. Along Paths II-V, the $O_2$ molecule directly dissociates onto the *ortho*-carbon-nitrogen site, *meta*-carbon-nitrogen site, *para*-carbon site and *meta*-carbon site from their corresponding physisorption sites of R states, respectively, by passing through the TS with the elongation of O-O bond and the formation of dangling C-O or N-O bonds. Similar to the physisorption of $O_2$ on the hollow site of the carbon ring, the $O_2$ molecule will first translate to the *meta*-carbon site by overcoming a barrier of 0.67 eV (see PS. 2. in SI). However, the relative energies of P states in Paths I-III show positive values of 2.58 eV, 1.31 eV and 1.11 eV while those in Paths IV and V are -0.37 eV and -0.28 eV, suggesting that the $O_2$ dissociation onto the *para*-nitrogen, *ortho*-carbon-nitrogen and *meta*-carbon-nitrogen atoms are endothermic and onto the *para*-carbon and *meta*-carbon atoms are exothermic.

**Table 2**
Calculated total energy ($E_t$), adsorption energy of two chemisorbed O atoms ($E_{ad} = E_{2O@C_3N} - 2E_o - E_{C_3N}$), Hirshfeld charges of two O atoms ($q_O$) and two O-bonded atoms ($q_C$ or $q_N$), distances between two dangling O atoms ($d_{OO}$), distances between two O-bonded atoms ($d_{NN}$, $d_{CN}$ or $d_{CC}$) of P states with two chemisorbed dangling O atoms on the carbon-nitrogen ring of $C_3N$ monolayer, and the imaginary frequency ($\omega$) of the related TS.

| Sites in P state | $E_t$/eV | $E_{ad}$/eV | $q_O$/e | $q_{C/N}$/e | $d_{OO}$/Å | $d_{NN/CN/CC}$/Å | $\omega$/cm$^{-1}$ |
|---|---|---|---|---|---|---|---|
| *para*-nitrogen (Path I) | -642.92 | -4.44 | -0.31<br>-0.31 | +0.11<br>+0.11 | 3.19 | 2.89 | 502 |
| *ortho*-carbon-nitrogen (Path II) | -644.17 | -5.70 | -0.30(C)<br>-0.30(N) | +0.05(C)<br>+0.05(N) | 2.51 | 1.85 | 782 |
| *meta*-carbon-nitrogen (Path III) | -644.38 | -5.90 | -0.33(C)<br>-0.30(N) | +0.07(C)<br>+0.11(N) | 3.06 | 2.55 | 406 |
| *para*-carbon (Path IV) | -645.85 | -7.37 | -0.33<br>-0.33 | +0.08<br>+0.08 | 3.19 | 2.89 | 458 |
| *meta*-carbon (Path V) | -645.76 | -7.28 | -0.32<br>-0.32 | +0.07<br>+0.07 | 2.92 | 2.50 | 301 |

The remark difference of relative energies of P states also implies the stability of two dangling bonds on the carbon-nitrogen ring generated from the O$_2$ dissociation process. Two dangling O atoms prefer to stably covalent to the atoms in the following order: *para*-carbon atoms > *meta*-carbon atoms > *meta*-carbon-nitrogen atoms > *ortho*-carbon-nitrogen atoms > *para*-nitrogen atoms, as indicated by the corresponding values of $E_{ad}$ listed in Tab. 1, -7.37 eV < -7.28 eV < -5.90 eV < -5.70 eV < -4.44 eV. The relatively lower values of $E_{ad}$ indicate the existence of stable configurations with two dangling bonds, despite that the O atoms are negatively charged (see the values of $q_O$ in Tab. 2) and repel each other (see the values of $d_{OO}$ and $d_{NN/CN/CC}$ in Tab. 2).

*3.4. Electronic band structure and DOS of the oxidized C$_3$N monolayer*

The oxidized structures with two chemisorbed O atoms also exhibit various band structures and DOS. The pristine C$_3$N monolayer is an indirect semiconductor with the bandgap of 0.39 eV, consistent with the value given by Zhou et al [31]. As shown in Fig. 6, the valence bands of C$_3$N monolayers with two O atoms at the *para*-carbon, *ortho*-carbon sites of the carbon-ring, and *para*-carbon site of the carbon-nitrogen ring cross the Fermi level. This indicates that these oxidized C$_3$N monolayers become metallic, due to the orbital coupling between two O atoms and bare C$_3$N monolayer at the valence band edge. While for O atoms at the *meta*-carbon site of the carbon ring and *para*-nitrogen site of the carbon-nitrogen ring, the orbital hybridization mainly occurs at lower energies far from the Fermi level and their band gaps are widened to 0.41 eV and 0.67 eV, respectively. The narrower band gaps of 0.06 eV, 0.28 eV and 0.18 eV can be seen for O atoms at the *ortho*-carbon-nitrogen site, *meta*-carbon-nitrogen site and *meta*-carbon site of the carbon-nitrogen ring, an indication of the enhanced electronic activity. These results demonstrate that the electronic properties of C$_3$N monolayer can be altered significantly due to the presence of two chemisorbed O atoms, from the metal to the semiconductor with the largest band gap of 0.67 eV.

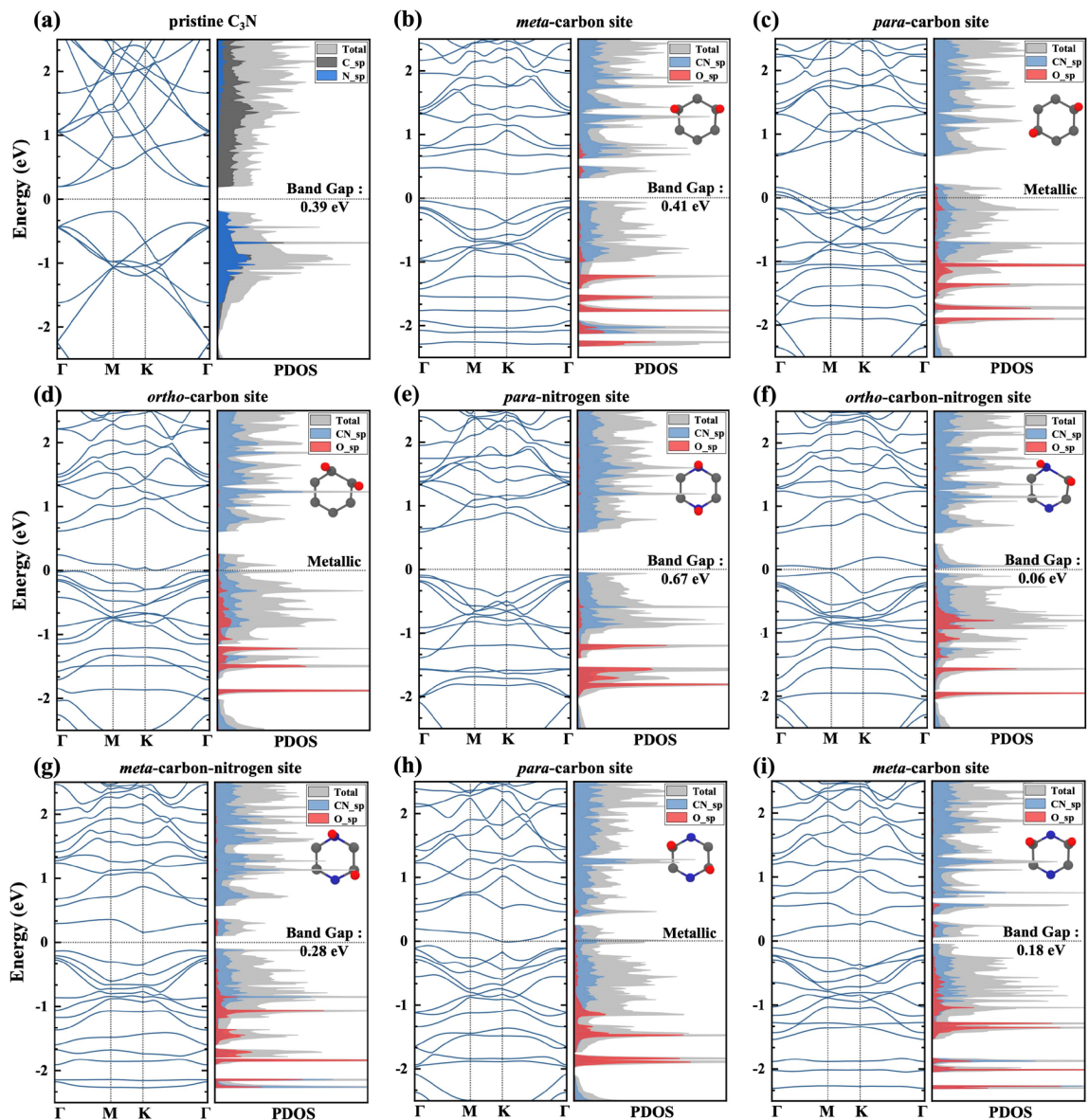

**Fig. 6.** The electronic band structure, DOS and projected density of states (PDOS) of the (a) pristine C$_3$N monolayer and eight products of O$_2$ dissociation including two O atoms on the carbon ring: (b) *meta*-carbon site, (c) *para*-carbon site and (d) *ortho*-carbon site, and on the carbon-nitrogen ring: (e) *para*-nitrogen site, (f) *ortho*-carbon-nitrogen site, (g) *meta*-carbon-nitrogen site, (h) *para*-carbon-nitrogen site and (i) *meta*-carbon site. The total DOS projected onto two O atoms and the bare C$_3$N monolayer are shown in red and blue areas, respectively. The inset only shows the carbon/carbon-nitrogen rings with two chemisorbed O atoms of the optimized configurations of products for clarity.

*3.5. Free energy for the oxygen dissociation*

To describe the oxygen dissociation at room temperature, we analyze the free energy change $\Delta G$ of TS and P state with reference to the initial state (O$_2$ + C$_3$N) along eight possible O$_2$ dissociation pathways. As shown in Tab. 3, $\Delta G_{TS}$ and $\Delta G_P$ at $T$ = 300 K increase by 0.1 ~ 0.2 eV compared to $\Delta E_{TS}$ and $\Delta E_P$ at $T = 0$ K, due to the difference of zero-point energy and that of entropy contribution (see PS. 3. in SI for $\Delta G_{TS}$ and $\Delta G_P$ at other temperatures). We can see among these dissociation pathways at $T = 300$ K, the two-step dissociation process occurring at the *ortho*-carbon of the carbon ring shows the lowest barrier of 0.90 eV and two chemisorbed O atoms at the *para*-carbon of the carbon ring shows the lowest free energy of -1.28 eV. This indicates that the most preferable dissociation pathway is still the two-step process and two chemisorbed O atoms at the *para*-carbon site of the carbon ring still corresponds to the most stable dissociation structure.

**Table 3**
Calculated free energy change $\Delta G$ of TS and P state with reference to the initial state ($O_2$ + $C_3N$) along eight possible $O_2$ dissociation pathways at the temperature $T = 300$ K. $\Delta E_{TS}$, $\Delta E_P$ are energy changes of TS and P state at $T = 0$ K without the zero-point energy contributions[a].

| | Carbon Ring | | | Carbon-Nitrogen Ring | | | | |
|---|---|---|---|---|---|---|---|---|
| | *meta*-carbon (Path I) | *para*-carbon (Path II) | *ortho*-carbon (Path III) | *para*-nitrogen (Path I) | *ortho*-carbon-nitrogen (Path II) | *meta*-carbon-nitrogen (Path III) | *para*-carbon (Path IV) | *meta*-carbon (Path V) |
| $\Delta E_{TS}$ | 1.46 | 1.19 | 0.75 (TS$_1$) 0.32 (M) 0.37 (TS$_2$) | 4.07 | 2.50 | 2.25 | 1.40 | 1.00 |
| $\Delta E_P$ | -1.28 | -1.49 | -1.33 | 2.22 | 0.97 | 0.76 | -0.70 | -0.62 |
| $\Delta G_{TS}$ | 1.63 | 1.34 | 0.90 (TS$_1$) 0.51 (M) 0.53 (TS$_2$) | 4.15 | 2.56 | 2.36 | 1.55 | 1.11 |
| $\Delta G_P$ | -1.08 | -1.28 | -1.11 | 2.39 | 1.17 | 0.94 | -0.51 | -0.43 |

[a]Values of $\Delta E_{TS}$, $\Delta E_P$ are not equal to the relative energies of TS and P state in Figs. 4 and 5 where the reference states are the corresponding physisorbed states. $\Delta G$ is calculated according to Eq. (3) given in Sec. 2.3. M denotes the intermediate state in a two-step dissociation process.

*3.6. Phase diagram of $O_2$ adsorption at experimental conditions*
To understand the thermodynamic stability of oxidized $C_3N$ structures at experimental conditions, we introduce the adsorption Gibbs free energy $\Delta G^{ad}(T, p_{O_2})$, which measures the Gibbs free energy of the oxidized state with respect to the reactant ($O_2$ + $C_3N$) at the temperature $T$ and $O_2$ partial pressure $p_{O_2}$. A more negative $\Delta G^{ad}$ indicates a more stable oxidized structure. According to the expression $\Delta G^{ad}(T, p_{O_2}) \approx E_{ad} - 2\Delta\mu_O(T, p_{O_2})$ (see details in Sec. 2.3.) deduced from *ab initio* atomistic thermodynamics, we can see that the formation of oxidized C₃N monolayer with two chemisorbed O atoms is accompanied by adsorbing an O₂ molecule onto the C₃N surface yielding an energy descend of $|E_{ad}|$, that is balanced by the free energy ascend of $|2\Delta\mu_O|$ due to the loss of an O₂ in the surrounding gas phase. $\Delta\mu_O$ is usually negative since that entropy contributions from translation and rotation are dominant and their values are positive.

Figure 7(a) plots the adsorption Gibbs free energy of an O₂, $\Delta G^{ad}$ versus the chemical potential of O, $\Delta\mu_O$, for eight possible oxidized structures. Despite the different values of $E_{ad}$, $\Delta G^{ad}$ all show the linear dependence on the $\Delta\mu_O$ with the same slope due to the same number of chemisorbed O atoms on the C₃N monolayer. At a certain $\Delta\mu_O$, $\Delta G^{ad}$ for the two chemisorbed O atoms at the *para*-carbon site of a carbon ring is lower than the values for other configurations, indicating that this configuration is the most favorable oxidized structure. We thus focus on this oxidized structure. It can be seen that $\Delta G^{ad} > 0$ for $\Delta\mu_O <$ -0.745 eV and a more negative $\Delta\mu_O$ results in a more positive $\Delta G^{ad}$. This means that when $\Delta\mu_O < -0.745$ eV, the O₂ molecule prefers to stay in the gas phase and a clean C₃N surface is obtained. While the oxidized structure with two O atoms adsorbed at the *para*-carbon site of a carbon ring is more stable when $\Delta\mu_O > -0.745$ eV as $\Delta G^{ad} < 0$. Thus, the value of $\Delta\mu_O = \frac{1}{2}E_{ad} = -0.745$ eV is a critical point which decides the existence of clean C₃N monolayer or the oxidized C₃N monolayer. This condition can be converted into a line in the $p_{O_2} - T$ phase diagram separating the phases of clean and oxidized C₃N monolayer, as shown in Fig. 7(b).

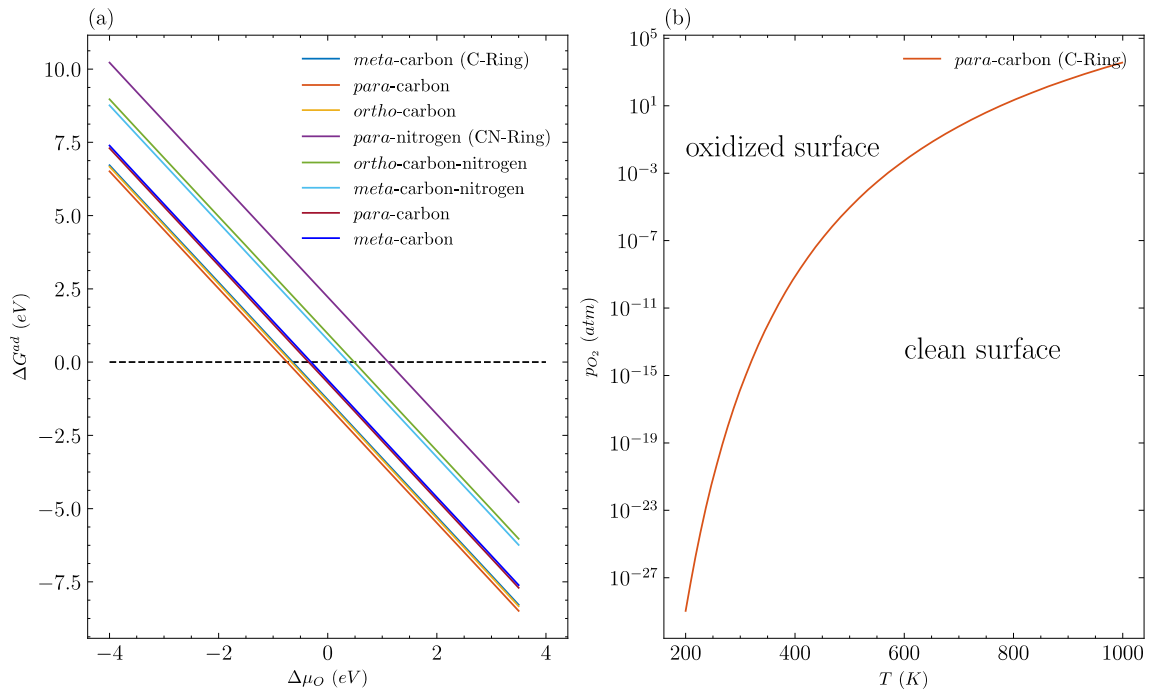

**Fig. 7.** (a) Calculated adsorption Gibbs free energy of an O₂ molecule for eight possible oxidized structures and (b) the surface phase diagram for the most stable product where two dangling O atoms are bonded to the *para*-carbon site of a carbon ring on the C₃N

monolayer. In the legend of (a), the former three lines represent the O$_2$ dissociation sites at the carbon ring and the remaining lines denote the sites at the carbon-nitrogen ring.

Whether the O$_2$ dissociation on the C$_3$N monolayer is feasible under realistic conditions depends on the height of barrier, which can be estimated by $\Delta G^{ad}$ using the value of $E_{ad}$ corresponding to the transition state. We focus on the two-step O$_2$ dissociation reaction since this process shows the lowest barrier of all eight possible reaction pathways, and choose the O$_2$ dissociation on the graphene for the comparison. The O$_2$ dissociation reaction is considered to be feasible when the barrier height is lower than 0.5 eV. This means that the critical value $\Delta G^{ad} \approx E_{ad} - 2\Delta\mu_O = 0.5$ eV. As $E_{ad} = 0.75$ eV for C$_3$N and 2.39 eV [40] for graphene, we obtain $\Delta\mu_O(T, p_{O_2}) = 0.125$ eV and 0.945 eV, respectively. As shown in Fig. 8, the minimal O$_2$ partial pressure $p_{O_2}$ needed for the feasible O$_2$ dissociation decreases with the increase of temperature $T$, indicating that lower O$_2$ partial pressure is sufficient to activate the O$_2$ dissociation on the C$_3$N or graphene at higher temperatures. However, at a certain temperature $T$, the O$_2$ is easier to be dissociated on the C$_3$N monolayer than on graphene since the value of $p_{O_2}$ is much smaller. At the atmospheric pressure of 1 *atm* and $T = 300$ K, the $p_{O_2} \sim 0.2$ *atm*, which is much smaller than the value of $p_{O_2}$ shown in Fig. 8, suggesting that the spontaneous O$_2$ dissociation on the C$_3$N or graphene under environmental conditions is very difficult.

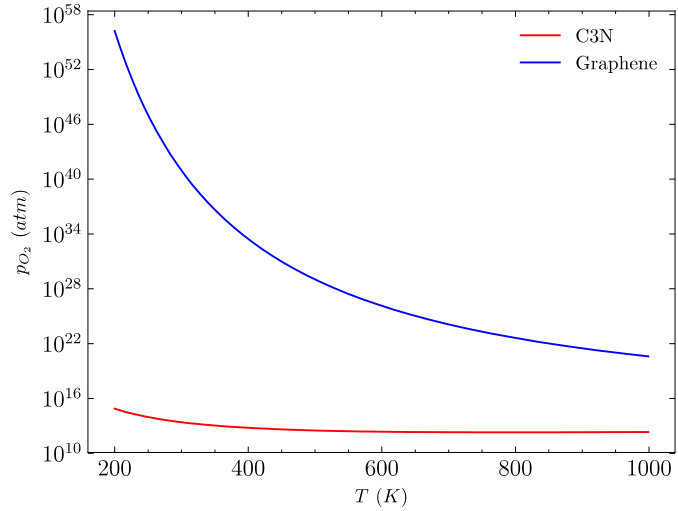

**Fig. 8.** The $p_{O_2} - T$ phase diagram of feasible occurrence of O$_2$ dissociation on the C$_3$N monolayer and graphene. The lines represent the minimal O$_2$ partial pressure needed for the feasible O$_2$ dissociation. They are plotted using the $\Delta\mu_O(T, p_{O_2}) = 0.125$ eV (C$_3$N monolayer, red line) and 0.945 eV (graphene, blue line), all corresponding to the barrier height of 0.5 eV.

## 4. Conclusions

In summary, we have performed the first-principles calculations to systematically investigate various adsorption sites of an oxygen molecule and an atomic oxygen, eight oxygen dissociation pathways and the associated oxidized structures on the pristine $C_3N$ monolayer. In comparison with the pristine graphene, the pristine $C_3N$ monolayer shows more $O_2$ physisorption sites and exhibits the stronger $O_2$ adsorption. The $O_2$ molecule can be adsorbed onto several sites at the carbon ring (*ortho*-, *meta*- and *para*-carbon atoms, hollow) or at the carbon-nitrogen ring (*para*- and *meta*-carbon, *ortho*- and *meta*-carbon-nitrogen, *para*-nitrogen atoms and hollow). The adsorption energies at these sites are close and over the range of 0.34 - 0.36 eV which are two times lower than that on the pristine graphene.

The $O_2$ dissociation pathways on the pristine $C_3N$ monolayer depends on the initial physisorbed sites and the oxidized structures, and the most preferable one is a two-step process involving an intermediate with the chemisorbed $O_2$ onto the *ortho*-carbon atoms of the carbon ring and the dissociation barrier is 1.10 eV, which is lower than about 2 eV for the similar process on the pristine graphene. Other $O_2$ dissociation pathways are direct routes where $O_2$ directly dissociates into two chemisorbed O atoms by forming two stable dangling C-O or N-O bonds, overcoming the barriers over the range of 1.3 - 4.4 eV. These results indicate that the pristine $C_3N$ monolayer is more susceptible to oxidation than the pristine graphene.

The presence of nitrogen atoms in the pristine $C_3N$ monolayer assists the formation of stable dangling C-O or N-O bonds, leading to the enriched oxidized structures after oxygen dissociation. For the dissociation state, two O atoms prefer to be stably chemisorbed onto the carbon or nitrogen atoms in the form of dangling C-O or N-O bonds. The oxidized structure with two dangling C-O bonds on the *para*-carbon atoms of the carbon ring shows the lowest adsorption energy of -8.16 eV and is more favorable than other structures. Moreover, the electronic properties can be altered significantly, from the metal to the semiconductor with the largest band gap of 0.67 eV.

The detailed description of $O_2$ dissociation and the oxidized structure on the pristine $C_3N$ monolayer can be generalized to a wide range of temperatures and pressures by analyzing the adsorption Gibbs free energy and surface phase diagram according to the *ab initio* atomistic thermodynamics. This provides insights into the tailoring of the surface chemical structures to manipulate the physical and chemical properties. For instances, the surface functionalization of the pristine $C_3N$ monolayer via the chemisorbed oxygen atoms is an effective method to tune the electronic and magnetic properties which sensitively depend on the surface oxygen coverage [74]. The dangling C-O bond on the pristine $C_3N$ monolayer can be utilized to enhance the adsorption of metal atoms on its surface to assist the single-atom catalysts [75]. It thus could be beneficial for the investigations on the catalysis and material engineering relative to the $C_3N$ or other carbon nitrides derivatives.

**Credit author statement**

L. Z. and Y. T. conceived, designed and guided the project. W. L. and Z. H. performed the calculations. W. L., Z. H., Z. Y. and H. J. analyzed the data. L. Z. and W. L. wrote the paper. All authors approved the final version of the manuscript.

**Declaration of competing interest**

The authors declare that they have no known competing financial interests or personal relationships that could have appeared to influence the work reported in this paper.

**Acknowledgments**

We acknowledge the valuable guidance given by Prof. Yi Gao, Prof. Zhongkang Han and PhD candidate Kai Zhang during the reply to reviewers. This work was supported by the National Natural Science Foundation of China [Nos: 11605151, 12075201, 11675138]; Natural Science Foundation of Jiangsu Province. [No: BK20201428]; Special Program for Applied Research on Supercomputation of the NSFC－Guangdong Joint Fund and Jiangsu Students' Innovation and Entrepreneurship Training Program (No. 202111117008Z).

# Supplementary Information for "Oxygen dissociation on the C₃N monolayer: A first-principles study"

Liang Zhao[1*], Wenjin Luo[1*], Zhijing Huang[1], Zihan Yan[1], Hui Jia[1], Wei Pei[1], and Yusong Tu[1†]

[1]College of Physical Science and Technology & Microelectronics Industry Research Institute, Yangzhou University, Jiangsu, 225009, China

*Authors contribute equally to this work
†Corresponding author: ystu@yzu.edu.cn

**PS. 1. O₂ translation pathway from the hollow site to the *meta*-carbon site on the carbon ring**

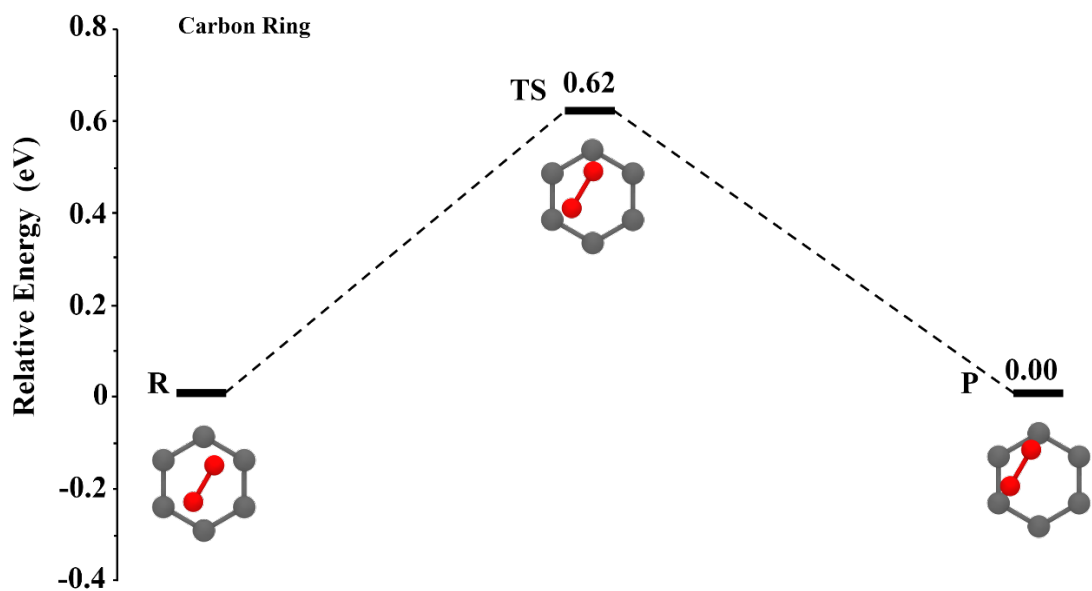

**Fig. S1.** O₂ translation pathway from the hollow site to the *meta*-carbon site on the carbon ring of the C₃N sheet and the optimized configurations. Notations: reactant (R), transition state (TS) and product (P). The energies of R state and P state are similar, and shifted to zero for an easy comparison. The number on the energy level denotes the relative energy.

**PS. 2. O₂ translation pathway from the hollow site to the *meta*-carbon site on the carbon-nitrogen ring**

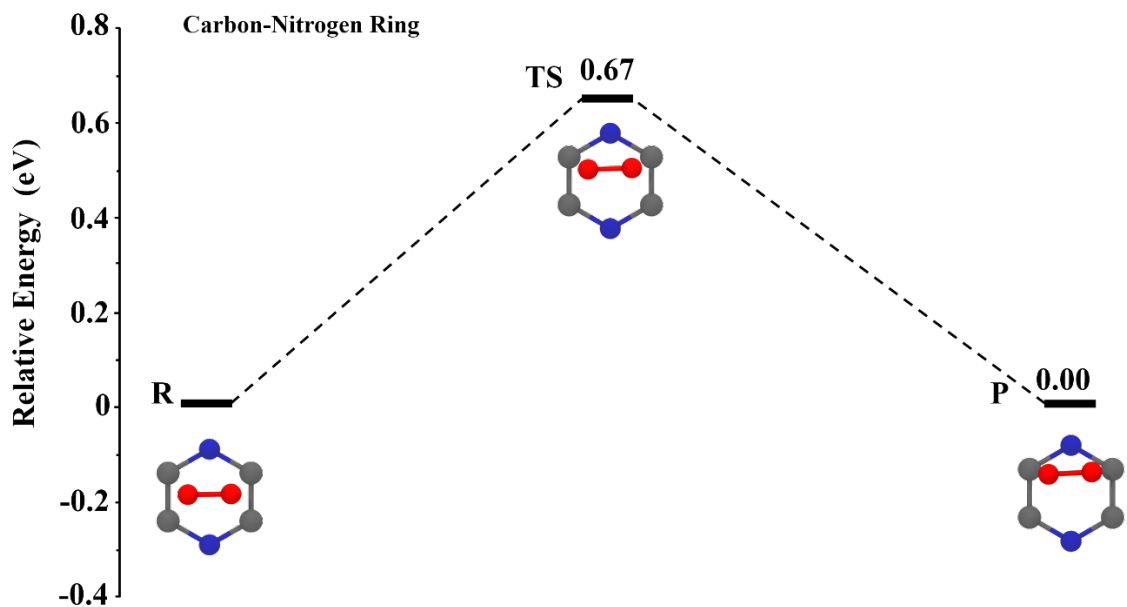

**Fig. S2.** $O_2$ translation pathway from the hollow site to the *meta*-carbon site on the carbon-nitrogen ring of the $C_3N$ sheet and the optimized configurations. Notations are same to those in Fig. S1.

**PS. 3. Temperature dependence of free energy change $\Delta G$ of the transition state and product state along eight possible $O_2$ dissociation pathways**

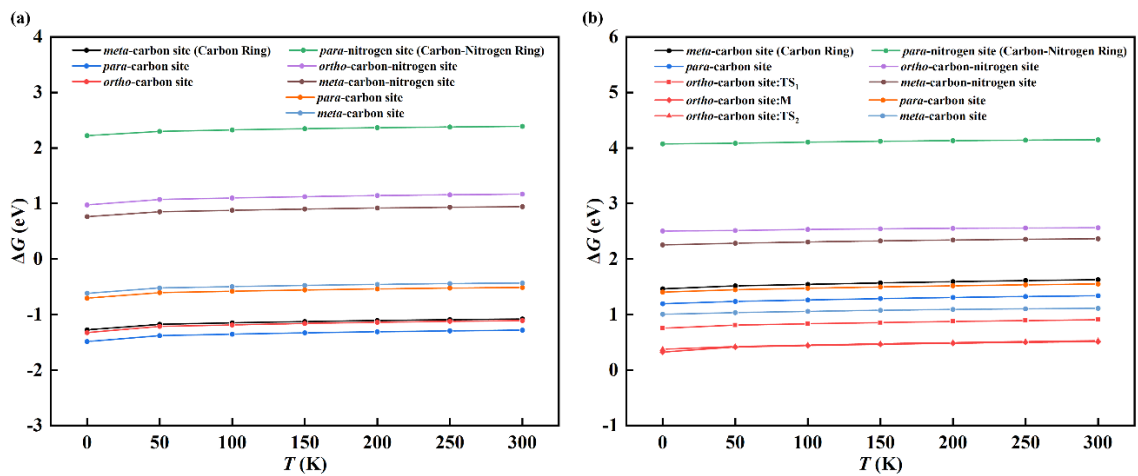

**Fig. S3.** The temperature dependence of free energy change $\Delta G$ of the (a) transition state and (b) product state along eight possible $O_2$ dissociation pathways. Notations: transition state (TS), intermediate (M) and product (P).